\documentclass[aps,pra,superscriptaddress, twocolumn,showpacs,floatfix]{revtex4-2}

\usepackage{amssymb,amsmath,amsfonts,mathrsfs,bm, bbm, graphicx, epsfig, epstopdf, mathtools}
\usepackage[version=3]{mhchem}
\usepackage{braket}
\usepackage{hyperref}
\usepackage{color}
\usepackage{xcolor}
\usepackage{adjustbox}
\usepackage{csquotes}
\usepackage{float}
\usepackage{colortbl}
\usepackage{siunitx}
\DeclareSIUnit{\nothing}{\relax}
\usepackage{makecell}
\usepackage[T1]{fontenc}
\usepackage{soul}

\usepackage{comment}
\usepackage{diagbox}
\usepackage{stfloats}

\date{\today}
\begin{document}
\title{Efficient Quantum Repeater with Single Atoms in Cavities}

\author{Yisheng Lei}
\thanks{Corresponding author}
\email{YishengLei2025@u.northwestern.edu}
\affiliation{Department of Electrical and Computer Engineering and Applied Physics Program, Northwestern University, Evanston, IL 60208, USA}

\begin{abstract}
Efficient quantum repeaters are needed to combat photon losses in fibers in future quantum networks. Single atom coupled with photonic cavity offers a great platform for photon-atom gate. Here I propose a quantum repeater scheme with efficient entanglement generation and entanglement swapping based on photon-atom gates. It can be implemented with various types of atomic systems and requires much less experimental complexity compared to other repeater protocols. With current available experimental techniques and reasonable improvements, high entanglement distribution rates can be achieved. A multiplexing configuration of 10 single atoms in cavities, secret key rates in order of a few Hz to 100s Hz can be achieved for communication distance of 1000~km. This proposal paves the way for the demonstration of an efficient entanglement distribution with quantum repeaters in the near future. 
\end{abstract}
\maketitle{}

\section{Introduction}
Quantum internet will be an enabling powerhouse for many quantum applications, such as distributed quantum computing, secured quantum communications, distributed quantum sensing, networked atomic clocks, etc \cite{kimble2008quantum, wehner2018quantum}. Since the initial proposal of entanglement generation between distant atomic systems \cite{cirac1997quantum}, significant progresses have been made for the past three decades. The concept of quantum repeater has been introduced to overcome photon loss and operational errors \cite{briegel1998quantum, azuma2023quantum}. Quantum repeater schemes based on quantum memory, such as DLCZ protocol (which refers as emissive quantum memory) has initialized the experimental efforts for developing long-distance quantum networks \cite{duan2001long}. Quantum repeaters based on absorptive quantum memories have triggered the experimental efforts for developing quantum memory devices with  high-efficiency, multimode, long memory time and on-demand properties \cite{simon2007quantum, lvovsky2009optical, lei2023quantum, gu2024hybrid}. To overcome the probabilistic nature of the entanglement generation and entanglement purification processes of quantum-memory based quantum repeaters, quantum repeaters with deterministic entanglement generation and error corrections have been proposed, for example by using quantum parity codes \cite{munro2012quantum, muralidharan2014ultrafast}, photonic graph states \cite{azuma2015all, buterakos2017deterministic, borregaard2020one}, GKP codes \cite{rozpkedek2021quantum}. Multiple schemes for entanglement generation with single atomic systems have been introduced, for instance single-photon interference protocols, such as single-photon \cite{cabrillo1999creation} and two-photon \cite{barrett2005efficient} entanglement schemes. There are also schemes based on two-photon interference, which has a maximum success probability of 50\% under the assumption of no photon loss and 100\% detector efficiency \cite{duan2003efficient}. All of these schemes relay on spontaneous photon emissions from an optically excited state which is stochastic process, in result the entanglement rates are low due to ineffective photon collections, especially for atoms with longer lifetime of the excited state \cite{sangouard2011quantum}. Bell state measurements of photonic qubits are typically used for entanglement swapping to extend entanglement distance \cite{zukowski1993event}, which suffer to be incomplete of 50\% success probability with simple linear optical setup \cite{grice2011arbitrarily, ewert20143}. Experimental demonstrations of elementary quantum networks have been achieved with various types of atomic systems, such as entanglement of multiple atomic ensembles separated by tens of kilometers away \cite{yu2020entanglement, liu2024creation}, entanglement distribution between NV centers in diamond \cite{pompili2021realization, stolk2024metropolitan}, SiV centers in diamond coupled with nanophotonic cavity \cite{knaut2024entanglement}, single cold atoms \cite{hofmann2012heralded, van2022entangling}, trapped ions \cite{krutyanskiy2023entanglement}, single rare-earth ions in crystals \cite{ruskuc2025multiplexed}, ensemble of rare-earth ions in solids \cite{lago2021telecom, liu2021heralded}, quantum dots in semiconductors \cite{delteil2016generation, stockill2017phase} and hot vapors \cite{li2021heralding}, however the entanglement rates are relatively low for actual quantum network applications, ranging from sub-Hz to kilo-Hz. 
Photon atom controlled-PHASE gate has been introduced \cite{duan2004scalable}, which has been used to demonstrate non-destructive photon detection \cite{niemietz2021nondestructive}, photon-atom gate \cite{reiserer2014quantum, tiecke2014nanophotonic}, photon-photon gate \cite{hacker2016photon}, atom-atom entanglement in a cavity \cite{welte2017cavity}, non-destructive and complete Bell state measurement (BSM) \cite{welte2021nondestructive}, atom-atom entanglement at distant nodes \cite{daiss2021quantum}. 
In this article, I propose an efficient quantum repeater scheme which consists of probabilistic entanglement generation and nearly deterministic entanglement swapping. The scheme is based on CNOT gate operations between single photons and single atoms in high-cooperativity cavities.

\section{Quantum repeater scheme}
The quantum repeater scheme is based on single photon sources (SPS) and single atoms in cavities. To distribute entanglement between two parties separated by a distance L, it is divided into n nodes with elementary link length $\frac{L}{2^{n-1}}$, and $2^{n-1}-1$ repeater stations, each of which has two repeater nodes. Entanglement generation are performed at every elementary link simultaneously and results are shared between the two nodes. Then, first level of entanglement swapping are carried out and results are feed forward to the neighboring stations; second level of entanglement swapping continue until $(n-1)^{th}$ level. The entanglement distribution cycle is complete. Entanglement purification can be followed.

\subsection{Photon-atom gate}

\begin{figure}[!h]
\centerline{\includegraphics[width=0.8\columnwidth]{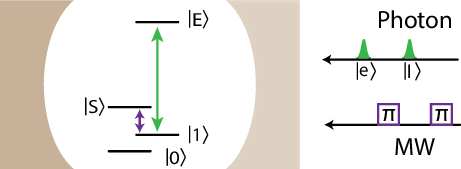}}
	\caption{(a) A single atom with two ground states $\ket{0}$ and $\ket{1}$, one shelving state $\ket{S}$ and one excited state $\ket{E}$. The optical transition between $\ket{1}$ and $\ket{E}$ is resonantly coupled with a single-sided cavity.}
	\label{Fig1}
\end{figure}

Here I describe a controlled-phase gate between a photonic qubit in time-bin basis and a single atom in a high-finesse cavity. If the photon is in early time-bin, denoted as $\ket{e}$, while as $\ket{l}$ if in late time-bin, separated by $\delta t$. The atom has two ground states $\ket{0}$ and $\ket{1}$, which serves as the qubit states. In addition, it has an optically excited state $\ket{E}$ and a shelving state $\ket{S}$. The cavity is a single-sided cavity, so photons sent through it always gets reflected. The resonance of the cavity is tuned to match the transition frequency between states $\ket{E}$ and $\ket{1}$, which is $f_0$. A photon of frequency $f_0$ encoded in time-bin basis is sent to the cavity. If the atom is in state $\ket{1}$, the photon gets reflected before entering the cavity, resulting in zero phase shift, while the atom is in state $\ket{0}$, the photon enters the cavity then gets reflected, resulting in a $\pi$ phase shift. After the early time-bin photon gets reflected and before the arrival of the late time-bin photon, a microwave $\pi$-pulse resonant with states between $\ket{1}$ and $\ket{S}$ is applied to the atom. After the late time-bin photon gets reflected, a microwave $\pi$-pulse is applied to the atom. If the atom is state $\ket{1}$, it undergoes one Rabi oscillation, so a $\pi$ phase shift between states $\ket{0}$ and $\ket{1}$ is created. (Please note that a microwave 3$\pi$-pulse should be applied to the atom based on Duan-Kimble scheme, instead this is a modified version.) The operations shown in Fig. \ref{Fig1} give the following:
\begin{equation}
\begin{split}
\begin{gathered}
     \ket{0}\ket{e} \rightarrow -\ket{0}\ket{e}; \\
     \ket{1}\ket{e} \rightarrow -\ket{1}\ket{e}; \\
     \ket{0}\ket{l} \rightarrow -\ket{0}\ket{l}; \\
     \ket{1}\ket{l} \rightarrow \ket{1}\ket{l}.
\end{gathered}
\end{split}
\end{equation}

This is a controlled phase gate between the photonic qubit and the atomic qubit. If the photonic qubit is represented in basis $\ket{\pm}_{p}$ = 1/$\sqrt{2}$($\ket{e}$ $\pm$ $\ket{l}$), the controlled-phase gate transforms to be a controlled-NOT (CNOT) gate with the atomic qubit as control qubit and the photonic qubit as target qubit, shown as the following:
\begin{equation}
\begin{split}
     \ket{0}\ket{+} \rightarrow -\ket{0}\ket{+}; \\
     \ket{1}\ket{+} \rightarrow -\ket{1}\ket{-}; \\
     \ket{0}\ket{-} \rightarrow -\ket{0}\ket{-}; \\
     \ket{1}\ket{-} \rightarrow -\ket{1}\ket{+}.
\end{split}
\end{equation}
Global phase can be ignored here. 
Cooperativity of the atom-cavity system is defined as $C = \frac{4g^2}{\kappa\Gamma}$, where $g$ is the coupling strength between the atom and the cavity, $\kappa$ is the decay rate of the cavity and $\Gamma$ is the decay rate of the atom's excited state. The success probability and fidelity of this CNOT gate requires high cooperativity (typically 100 - 1000), low photon loss, perfect photon-cavity mode matching, high value of $\kappa T_p$, where $t_p$ is the duration of the photon wavepacket \cite{duan2004scalable, zhan2020deterministic, borregaard2020one, nagib2024robust}. Other schemes of photon-atom CNOT gate are discussed in Appendix. \ref{CNOTO}. 

\subsection{Entanglement generation}

\begin{figure}[!h]
\centerline{\includegraphics[width=0.95\columnwidth]{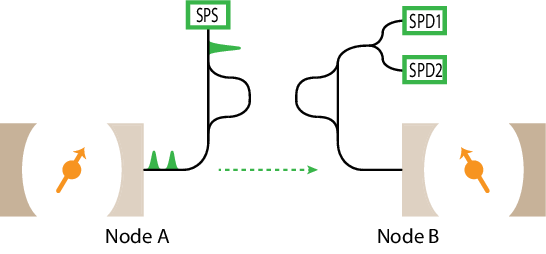}}
	\caption{(a) At Node A, there are a single photon source (SPS) and a single atom coupled with a cavity. At Node B, there are two single photon detectors (SPD) and a single atom coupled with a cavity.}
	\label{Fig2}
\end{figure}

Here I describe how to generate an entangled state between two distant single atoms in cavities by sending a single photon to both of the cavities. A deterministic or near-deterministic single photon source (SPS) is located at the Node A. The single photons pass through an unbalanced Mach-Zehnder interferometer, which leaves them in state $\ket{+}_p$, then they are directed to the cavity. The single atoms are initially prepared in the ground state $\ket{0}$, then a microwave $\pi/2$-pulse is apllied to the single atoms, which leave them in a superposition state $\ket{+}_A$ = $\frac{1}{\sqrt{2}}(\ket{0}+\ket{1})$. A CNOT operation between the photon and the atom is performed, then the photon is sent to the Node B. A CNOT operation is performed between the photon and the atom, then the photon is routed to an unbalanced Mach-Zehnder interferometer. Single photon detection (SPD) is performed. 
When the photon is reflected from the cavity at the Node A, the state of the photon and the atom becomes $-\frac{1}{\sqrt{2}} \left( \ket{0}_{A}\ket{+}_{p} + \ket{1}_{A}\ket{-}_{p} \right) $ (global phase can be ignored here.). After reflection from the cavity at the Node B, assuming no photon loss, the state of the two single atoms and the photon becomes 
\begin{equation}
        \ket{A, B, p} = \frac{1}{\sqrt{2}}
        \left[
        \begin{array}{cc}
            \frac{1}{\sqrt{2}} \left( \ket{0}_{A}\ket{0}_{B} + \ket{1}_{A}\ket{1}_{B}\right) \ket{+}_p \\
            + \frac{1}{\sqrt{2}} \left(\ket{0}_{A}\ket{1}_{B} + \ket{1}_{A}\ket{0}_{B}\right) \ket{-}_p
        \end{array}
        \right].
\label{eqa3}
\end{equation}


If the SPD measurement outcome is $\ket{+}$, the two distant single atoms are entangled with $\ket{\Phi^+_{AB}}$; while the measurement outcome is $\ket{-}$, the two single atoms are entangle with $\ket{\Psi^+_{AB}}$, a bit flip operation is carried out to the atom at the Node B, which is a microwave $\pi$-pulse. At the end, the two distant single atoms are entangled with $\ket{\Phi^+_{AB}}$. If the photon is in a vacuum state initially, the two atoms are not disturbed. If the photon gets reflected successfully from both of the cavities but the photon is not detected, the two atoms are either in state $\frac{1}{\sqrt{2}} \left( \ket{0}_{A}\ket{0}_{B} + \ket{1}_{A}\ket{1}_{B}\right)$ or $\frac{1}{\sqrt{2}} \left(\ket{0}_{A}\ket{1}_{B} + \ket{1}_{A}\ket{0}_{B}\right)$. In this case, if another photon is sent and same operations are performed, assuming the second photon is not lost, the state of the two single atoms and the second photon becomes
\begin{equation}
        \ket{A, B, p_2} = \frac{1}{\sqrt{2}}
        \left[
        \begin{array}{cc}
            \frac{1}{\sqrt{2}} \left( \ket{0}_{A}\ket{0}_{B} + \ket{1}_{A}\ket{1}_{B}\right) \ket{+}_{p_2} \\
            + \frac{1}{\sqrt{2}} \left(\ket{0}_{A}\ket{1}_{B} + \ket{1}_{A}\ket{0}_{B}\right) \ket{-}_{p_2}
        \end{array}
        \right].
\end{equation}
If the first photon is reflected successfully from the cavity A but not from the cavity B, the atom A is in a mixture state of $-\ket{0}_A$ and $-\ket{1}_A$ (global phase can be ignored here.) and the atom B is in state $\ket{+}_B$. In this case, the two atoms are in a mixed state.
As we can see from the analysis above, only one single photon can be sent for one attempt of entanglement generation. This entanglement generation scheme is fundamentally different from other protocols, since it doesn't require atoms to decay. 

\subsection{Entanglement swapping}
Two elementary links are entangled with $\ket{\Phi^+_{AB}}$ and $\ket{\Phi^+_{CD}}$. Here I describe how to perform complete Bell state measurement between the single atoms at the Node B and C by sending two photons sequentially to both of the cavities. The first photon with state $\ket{+}$ is sent to the cavity at the Node C first, CNOT gate operation is performed between the photon and the atom, the photon is routed to the cavity at the Node B, CNOT gate operation is performed between the photon and the atom. At the end, the photon is routed to an unbalanced Mach-Zehnder interferometer. Single photon detection (SPD) is performed. If the measurement outcome is $\ket{+}$, the two single atoms are in the same state, which is $\ket{0}_B\ket{0}_C$ or $\ket{1}_B\ket{1}_C$, which means they are in $\ket{\Phi^{\pm}}_{BC}$; while the meaurement outcome is $\ket{-}$, the two single atoms are in opposite state, which is $\ket{0}_B\ket{1}_C$ or $\ket{1}_B\ket{0}_C$, which means they are in $\ket{\Psi^{\pm}}_{BC}$. This process is known as quantum parity checking. After successfully registered the first photon, a microwave $\pi/2$-pulse is applied to both the single atoms at the Node B and C, which transforms $\ket{0}$ into $\frac{1}{\sqrt{2}}$($\ket{0} + \ket{1}$) and $\ket{1}$ into $\frac{1}{\sqrt{2}}$($\ket{0} - \ket{1}$). Here I show an example, when the two single atoms are in state $\Phi^+_{BC}$, after $\pi/2$ rotations,


\begin{figure*}[ht]
\centerline{\includegraphics[width=1.85\columnwidth]{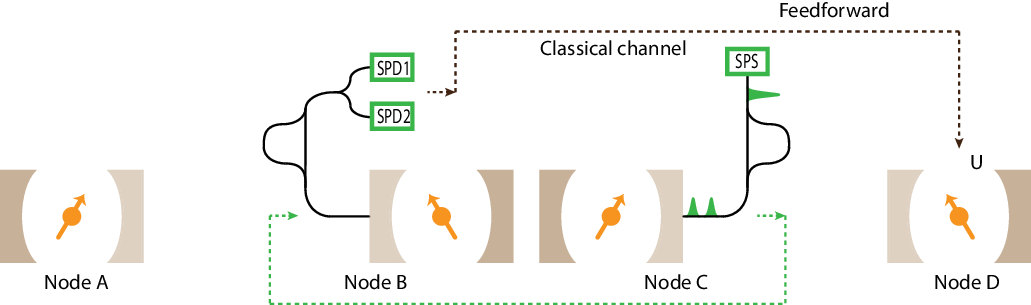}}
	\caption{(a) BSM between Node B and Node C are performed. Measurement results are sent to Node D, and unitary operations are applied to the atomic qubit accordingly.}
	\label{Fig3}
\end{figure*}


\begin{equation}
\begin{split}
\begin{gathered}
     \frac{1}{\sqrt{2}} \left( \ket{0}_{B}\ket{0}_{C} + \ket{1}_{B}\ket{1}_{C}\right) \rightarrow \\ 
    \frac{1}{\sqrt{2}} \left(\ket{0}_B + \ket{1}_B \right) \otimes \frac{1}{\sqrt{2}} \left(\ket{0}_C + \ket{1}_C \right) \\
    + \frac{1}{\sqrt{2}} \left(\ket{0}_B - \ket{1}_B \right) \otimes \frac{1}{\sqrt{2}} \left(\ket{0}_C - \ket{1}_C \right) \\
    \Leftrightarrow \frac{1}{\sqrt{2}} \left( \ket{0}_{B}\ket{0}_{C} + \ket{1}_{B}\ket{1}_{C}\right).
\end{gathered}
\end{split}
\end{equation}
After calculating all possible cases, we can see that $\Phi^+$ and $\Psi^-$ stay unchanged; while $\Phi^-$ changes into $\Psi^+$ and vice versa. The second photon is sent to the cavity at the Node C, and the same operations are performed until one photon is successfully registered. The SPD measurement outcomes and BSM results are listed in Table. \ref{Table1}.
\begin{table}[!h]
\begin{center}
\begin{tabular}{ c c c }
\hline\hline
 First photon & Second photon & BSM \\
 \hline
 + & + & $\Phi^+$ \\
 \hline
 + & - & $\Phi^-$ \\
 \hline
 - & + & $\Psi^+$ \\
 \hline
 - & - & $\Psi^-$ \\
 \hline
\end{tabular}
\caption{Truth table for BSM}
\label{Table1}
\end{center}
\end{table}
By default, state of the atom at the Node B is teleported to the atom at the Node D, and the atomic state at the Node B is denoted as $\left( \alpha \ket{0_B} + \beta \ket{1_B} \right)$. Unitary operation to the atom at the Node D is needed based on the BSM. Here entanglement swapping can be regarded as quantum teleportation from the Node B to the Node D. Since the Node A and B are initially entangled, after quantum teleportation, the Node A and D are entangled.
\begin{equation}
    \begin{split}
        \ket{\Phi^+_{BC}} \otimes \left( \alpha \ket{0}_D + \beta \ket{1}_D \right) \\
        \ket{\Phi^-_{BC}} \otimes \left( \alpha \ket{0}_D - \beta \ket{1}_D \right) \\
        \ket{\Psi^+_{BC}} \otimes \left( \alpha \ket{1}_D + \beta \ket{0}_D \right) \\
        \ket{\Psi^-_{BC}} \otimes \left( \alpha \ket{1}_D - \beta \ket{0}_D \right)
    \end{split}
\end{equation}
At the end, the atoms at the Node A and D are entangled with $\ket{\Phi^+_{AD}}$. Since the Node B and C are in the same repeater station, photon transmission loss are negligible. Single photons can be sent sequentially (here it is limited to 5 for the following analysis) which are also associated with CNOT gate operations, to make sure two photons are registered at single photon detectors. The success probability of entanglement swapping can be nearly 100\%. As we can seen, this BSM are non-destructive and complete, unlike BSM for photonic qubits. Other methods of complete BSM are discussed in Appendix. \ref{BSMO}. 

\subsection{Multiplexing}

\begin{figure}[!h]
\centerline{\includegraphics[width=0.95\columnwidth]{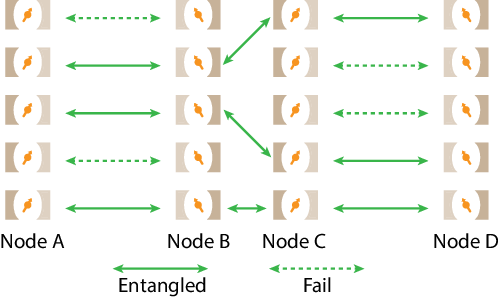}}
	\caption{Entanglement swapping can be performed between any neighboring entangled pairs. Here is an example of multiplexing with five atomic systems at each node.}
	\label{Fig4}
\end{figure}

Multiplexing technique can be used to greatly enhance entanglement distribution rates \cite{collins2007multiplexed, gu2024hybrid}. At each repeater node, multiple single atoms coupled with photonic cavities can be achieved with current laboratory technologies. The single atoms are not necessary in separate cavities, instead they can be placed in the same cavity as long as each single atom can be addressed separately, for example atom array. The entanglement generation processes at the elementary links have a success probability $p_0$ and the multiplexing number is $n_m$. The average number of entangled states generated by one cycle can be estimated as,

\begin{equation}
\begin{split}
\begin{gathered}
    N_{avg} = \Biggl\{ n_m\times\left(p_0^{n_m}\right)^{2^{n-1}} +  \\  
    \sum_{i=1}^{n_m-1} i \times \left[
    \begin{array}{cc}
    \sum_{k=1}^{2^{n-1}} {{2^{n-1}}\choose{k}} \\
    \times \left[{{n_m}\choose{i}}p_0^i \left(1-p_0\right)^{n_m-i} \right]^k \\
    \times \left[ \sum_{j=i+1}^{n_m} {{n_m}\choose{j}}p_0^j \left(1-p_0\right)^{n_m-j} \right]^{2^{n-1}-k} 
    \end{array} \right] \Biggr\} \\
    \times \left(1 - \left(1 - \eta_d\eta_s\eta_c^2p_{CN}^2 \right)^{n_s} \right)^{2(2^{n-1}-1)},
\end{gathered}
\end{split}
\end{equation}

where $\left(p_0^{n_m}\right)^{2^{n-1}}$ is the probability of $n_m$ number of entangled states between the two end nodes, and ${{n_m}\choose{i}}p_0^i \left(1-p_0\right)^{n_m-i}$ is the probability of $i$ number of entangled states at one elementary link and $\sum_{j=i+1}^{n_m} {{n_m}\choose{j}}p_0^j \left(1-p_0\right)^{n_m-j}$ is the probability of $\geq i$ number of entangled states at one elementary link. 
to ensure $i$ number of entangled pairs between two end nodes, one and only one elementary link has $i$ number of entangled pairs, the rest of elementary links ($2^{n-1}-1$) have $\geq i$ entangled states. $\left(1 - \left(1 - \eta_d\eta_s\eta_c^2p_{CN}^2 \right)^{n_s} \right)^2$ is the success probability of one entanglement swapping and ${(2^{n-1}-1)}$ is the total number of entanglement swapping at all stages.

\subsection{Repeater performance}
Single photon source with quantum efficiency above 70\% have been reported \cite{ding2025high}, and deterministic single photon source and on-demand single photons with high heralding efficiency are under active development \cite{ornelas2020demand}. Atom-cavity coupling efficiency has reached 99\% in multiple experimental systems \cite{bhaskar2020experimental, najer2019gated}. Coupling efficiency between a waveguide or cavity mode to a single mode fiber has reached 97\% \cite{tiecke2015efficient}. Single photon detectors with efficiency above 99\% have been reported \cite{chang2021detecting, reddy2020superconducting}. Coherence time of 6 hours have been demonstrated with nuclear spin of Eu$^{3+}$ ion in YSO crystal \cite{zhong2015optically}, as well as coherence time of 23 milliseconds with electronic spin of Er$^{3+}$ ions in CaWO$_4$ crystal \cite{le2021twenty}, and coherence time of 1s with nuclear spin of Er$^{3+}$ ion in YSO crystal \cite{ranvcic2018coherence}. Coherence time of more than one hour has been observed in trapped ions \cite{wang2021single}. Coherence time over one minute has been demonstrated with $\prescript{13}{}{\text{C}}$ nuclear spin coupled with NV center in diamond \cite{bradley2019ten}. $\prescript{29}{}{\text{Si}}$ nuclear spin in SiV center in diamond has shown coherence time over 2s \cite{stas2022robust}. High fidelity single qubit gate has been shown with various types of atomic systems \cite{bartling2025universal, ma2023high}.

\begin{table}[!h]
\begin{center}
\begin{tabular}{ c c c }
\hline\hline
 Parameters & Symbols & Values   \\
 \hline
 Qubit coherence time & $t_{coh}$ & 10s\\
 \hline
 Single photon efficiency & $\eta_s$ & 0.8 \\
 \hline
 CNOT gate time & $t_{CN}$ & 10$\mu s$ \\
 \hline
 SPD efficiency & $\eta_d$ & 0.99\\
 \hline
 Fiber-cavity coupling efficiency & $\eta_c$ & 0.99 \\
 \hline
 CNOT gate success probability & $p_{CN}$ & 0.99 \\
 \hline
 Fiber attenuation constant & $L_{att}$ & 22km \\
 \hline
 CNOT gate error rate & $\epsilon_{CN}$ & $10^{-4}$ \\
 \hline
 
\end{tabular}
\end{center}
\caption{List of parameters used for simulation in Fig. \ref{Fig5}.}
\label{Parameters}
\end{table}

The total time to complete one cycle of entanglement distribution can be estimated as,
\begin{equation}
    \begin{split}
    \begin{gathered}
         T_{tot} = 2t_{CN} + \frac{2L}{2^{n-1}c} \\
        + \sum_{i=0}^{n-2} \left( \frac{4t_{CN}}{\eta_d^2\eta_s^2\eta_c^4p_{CN}^4} + \frac{2^iL}{2^{n-1}c}\right),
    \end{gathered}
    \end{split}
\end{equation}
where $2t_{CN} + \frac{2L}{2^{n-1}c}$ is time required for entanglement generation and feed forward at the elementary links, and $\sum_{i=0}^{n-2} \left( \frac{4t_{CN}}{\eta_d^2\eta_s^2\eta_c^4p_{CN}^4} + \frac{2^iL}{2^{n-1}c}\right)$  is the time required for entanglement swapping and feed forward for (n-1) number of stages. Time consumed for feed-forward operation accounts for the majority of the total time. SPD time and time for single qubit gate operation during entanglement swapping are negligible. 

\begin{equation}
    p_0 = p_{CZ}^2\eta_c^2\eta_d \eta_s e^{-\frac{L}{2^{n-1}L_{att}}},
\end{equation}

Photonic qubit errors during transmission cause negligible effects to the fidelity of the entangled states, since photonic qubit in time-bin basis are robust against environmental noises. The main error sources are CNOT gate operations during entanglement generation and entanglement swapping. As the entanglement generation processes start, the qubit fidelity and the CNOT gate operation fidelity will drop due to qubit decoherence. The fidelity of the qubits with respect to time,
\begin{equation}
    \mathcal{F}_a = e^{-\frac{t}{t_{coh}}},
\end{equation}
and the fidelity of CNOT gate operation with respect to time,
\begin{equation}
    \mathcal{F}_{CN} = (1 - \epsilon_{CN})e^{-\frac{t}{t_{coh}}},
\end{equation}
The average fidelity of the entangled pairs can be estimated as,


\begin{equation}
    \begin{split}
    \begin{gathered}
        \mathcal{F}_{avg} = \left( 1 - \epsilon_{CN} \right) ^{2^{n}} \\ 
        \times\left( (1 - \epsilon_{CN})e^{-\frac{2t_{CZ} + \frac{2L}{2^{n-1}c}}{t_{coh}}} \right)^{\frac{4}{\eta_d^2\eta_s^2\eta_c^4p_{CN}^4}(2^{n-2})} \\
        \times \prod_{k=2}^{n-1}\left( (1 - \epsilon_{CN})e^{-\frac{t_k}{t_{coh}}} \right)^{\frac{4}{\eta_d^2\eta_s^2\eta_c^4p_{CN}^4}(2^{n-k-1})} \\
        \times e^{-\frac{T_{tot}}{t_{coh}}}
    \end{gathered}
    \end{split}
\end{equation}

\begin{equation}
    \begin{split}
    \begin{gathered}
       t_k = 2t_{CZ} + \frac{2L}{2^{n-1}c} \\
        + \sum_{i=0}^{k-2} \left( \frac{4t_{CN}}{\eta_d^2\eta_s^2\eta_c^4p_{CN}^4} + \frac{2^iL}{2^{n-1}c}\right)
        \end{gathered}
    \end{split}
\end{equation}

the first term is the fidelity after entanglement generations at the elementary links, where $2^{n}$ is the total number of CNOT gate operations at this step; the second term is the fidelity drop after first layer of entanglement swapping, $\frac{4}{\eta_d^2\eta_s^2\eta_c^4p_{CN}^4}(2^{n-2})$ is the total number of CNOT gate operations at this step; the third term is the entanglement drop after each layer of entanglement swapping from $2^{nd}$ layer to $(n-1)^{th}$ layer, where $\frac{4}{\eta_d^2\eta_s^2\eta_c^4p_{CN}^4}(2^{n-k-1})$ is the number of CNOT gate operations at $k^{th}$ layer; the fourth term is the qubit fidelity drop after the whole cycle.

Some of photons from the single-photon sources are in vacuum states since the single-photon sources are not required to be deterministic, and some photons are lost during transmission at the elementary links, so the qubit errors are slightly less than an actual CNOT gate with photon present, since there is no actual interaction between the photon and the atom in the cavity. In summary, I take the estimation of fidelity with lower bound, which makes the simulated results of secret key rates more convincing. The total number of CNOT gate operations can be calculated as,
\begin{equation}
    n_{CN} = 2^{n} + \frac{4}{\eta_d^2\eta_s^2\eta_c^4p_{CN}^4}(2^{n-1}-1),
\end{equation}
where $2^{n}$ is the total number of CNOT gate operations during entanglement generation processes at elementary links, and $\frac{4}{\eta_d^2\eta_s^2\eta_c^4p_{CN}^4}(2^{n-1}-1)$ is the total number of CNOT gate operations during entanglement swapping at multiple stages. Single qubit gate operation errors are negligible, since sufficiently high fidelity of single qubit gate operations have been achieved experimentally in multiple types of atomic systems. 

In this article, I consider quantum key distribution with six-state protocol \cite{bruss1998optimal, scarani2009security}. Its asymptotic key fraction is given by

\begin{equation}
    f = max \left\{ \left(1 - Q \right) \left[1 - h \left(\frac{1 - 3Q/2}{1 - Q} \right) \right] - h \left( Q \right), 0 \right\},
\end{equation}

where $h(x) = -x \text{log}_2 x - (1 - x)\text{log}_2(1-x)$ is the entropy function, Q is the qubit error rate.
\begin{equation}
    Q = 1 - \mathcal{F}_{avg}.
\end{equation}

Secret key rate is given by,
\begin{equation}
    R = \frac{1}{T_{tot}}N_{avg}f.
\end{equation}

\subsubsection{ L = 1000km}


\begin{figure*}[!ht]
\centerline{\includegraphics[width=1.8\columnwidth]{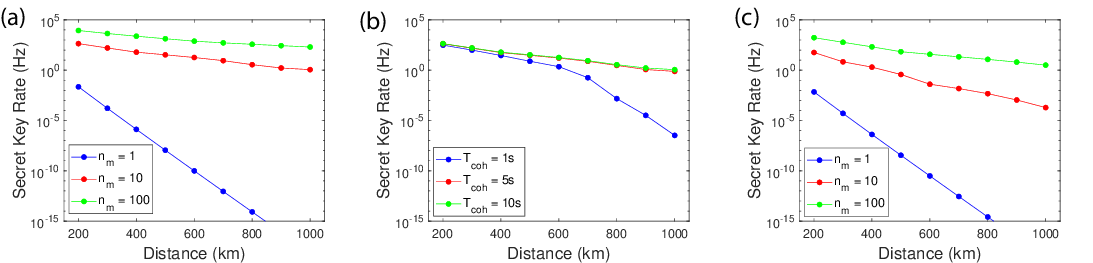}}
	\caption{Simulated secret key rates for communication distance between 200km and 1000km: (a) the parameters used for this simulation is listed at the Table. \ref{Parameters}; (b) $n_m$ = 10 for this simulation; (c) $p_{CN}$, $\eta_{d}$ and $\eta_{c}$ are 0.95 instead of 0.99.}
	\label{Fig5}
\end{figure*}

\begin{figure*}[!ht]
\centerline{\includegraphics[width=1.5\columnwidth]{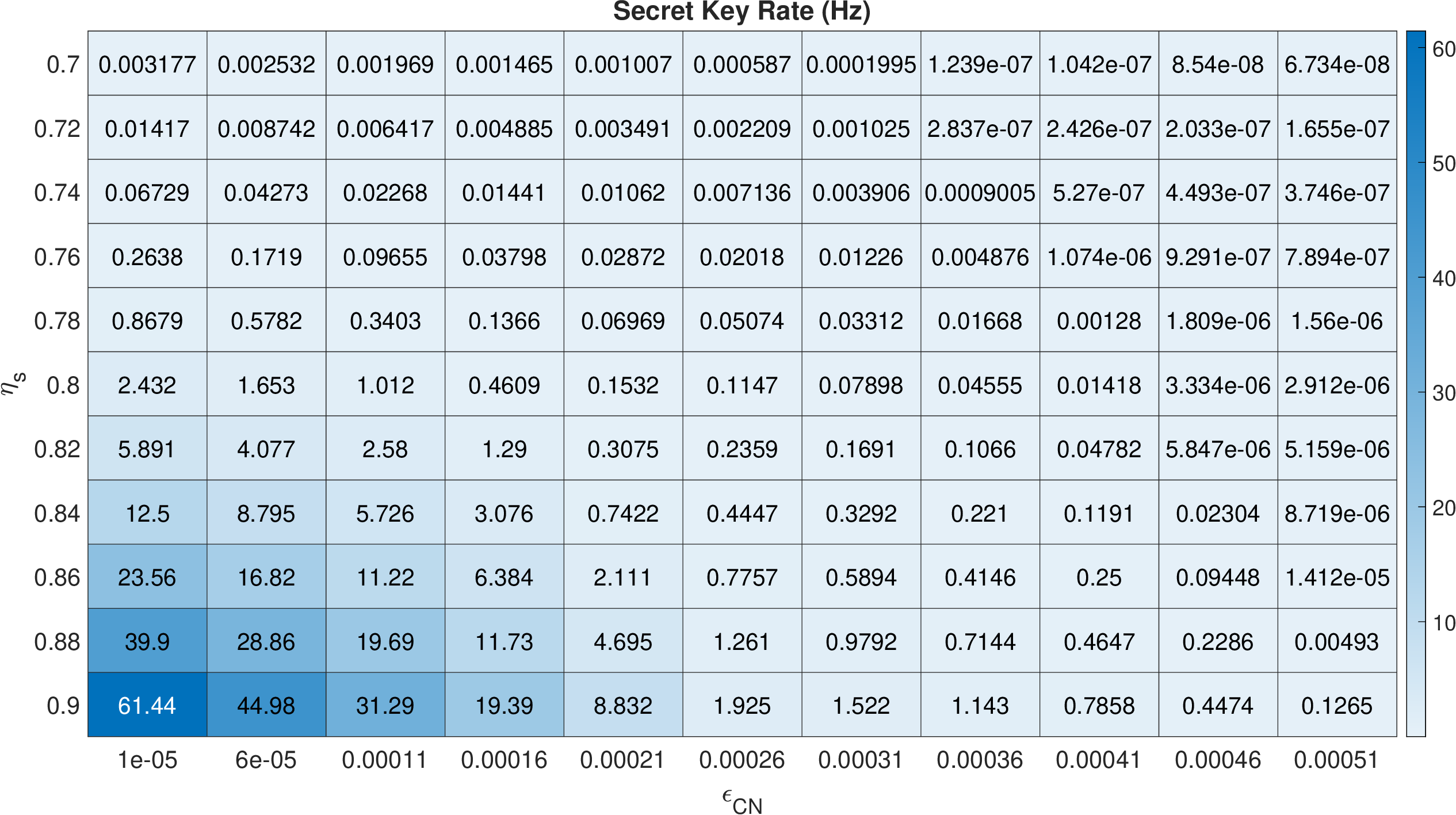}}
	\caption{Simulated secret key rates for communication distance of 1000km with CNOT gate error rates ranging from $10^{-5}$ to $5.1\times10^{-4}$ and single photon source efficiency between 0.7 and 0.9, $n_m$ = 10.}
	\label{Fig6}
\end{figure*}

\begin{figure*}[!ht]
\centerline{\includegraphics[width=1.5\columnwidth]{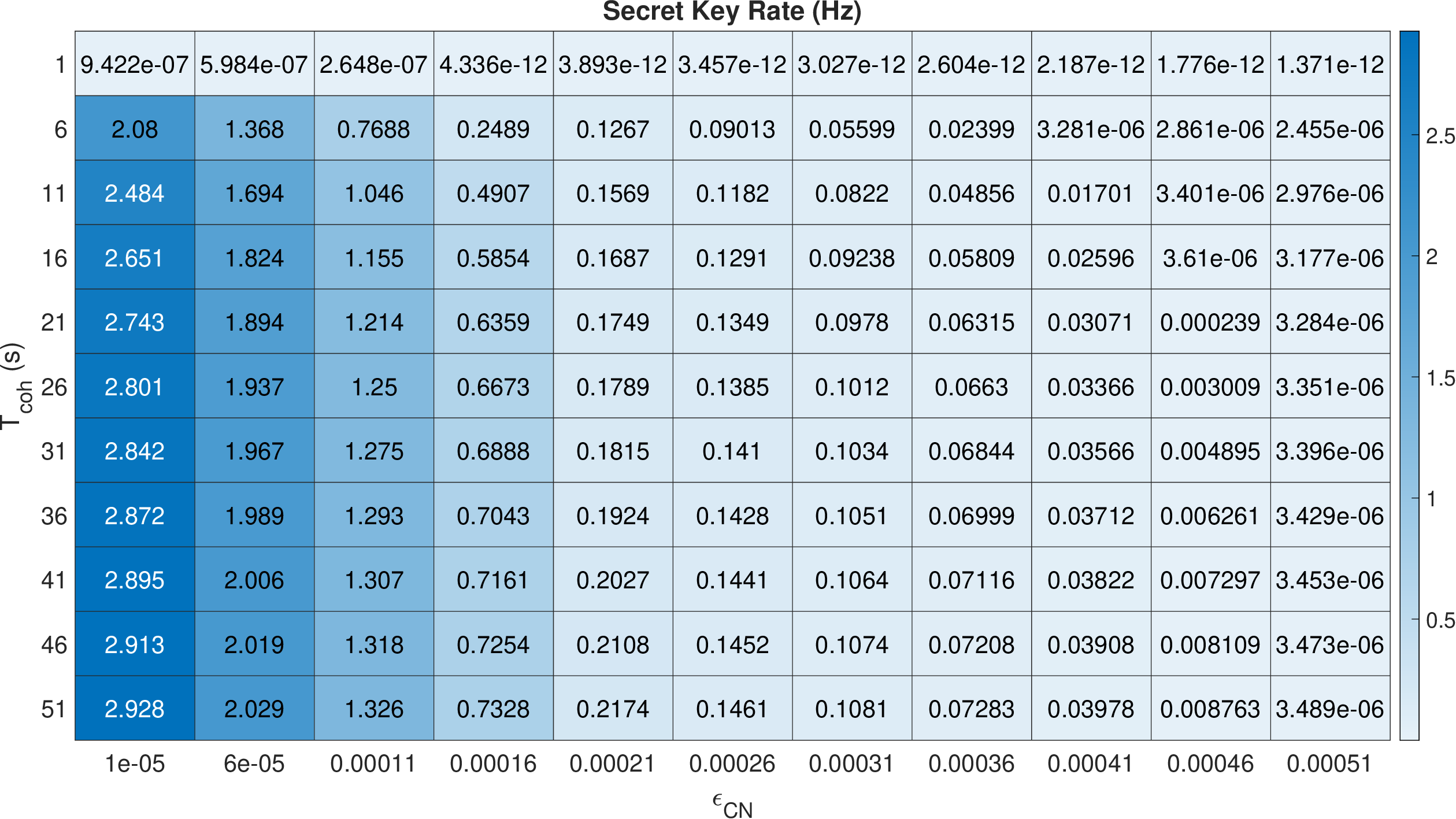}}
	\caption{(a) Simulated secret key rates for communication distance of 1000km with CNOT gate error rates ranging from $10^{-5}$ to $5.1\times10^{-4}$ and coherence time of the qubit between 1s and 51s, $n_m$ = 10.}
	\label{Fig7}
\end{figure*}

\twocolumngrid

Here I simulate the maximum secret key rates for communication distance of 1000km by finding the optimal repeater parameters such as n. By using the parameters listed in the Table. \ref{Parameters}, the results are shown in Fig. \ref{Fig5}(a). By keeping $n_m$ = 10, the results are listed in Fig. \ref{Fig5}(b) with different qubit coherence time. By changing $p_{CN}$, $\eta_{d}$ and $\eta_{c}$ to be 0.95, the results are listed in Fig. \ref{Fig5}(c), and the secret key rates drop by one or two orders of magnitude compared with the results in Fig. \ref{Fig5}(a). Secret key rates with different combination of single photon source efficiency and CNOT gate error rate are listed in Fig. \ref{Fig6}, with $n_m$ = 10, and some of the simulated optimal repeater parameters of n are listed at Table. \ref{Optimaln}. Other simulation results with different $n_m$ are listed in Appendix. \ref{RP}. Secret key rates with different combination of qubit coherence time and CNOT gate error rate are listed in Fig. \ref{Fig7}, with $n_m$ = 10. Other simulation results with different $n_m$ are listed in Appendix. \ref{RP}.

\begin{table}[!h]
\begin{center}
\begin{tabular}{ |c |c |c |c | }
\hline
\diagbox{$\eta_s$}{$\epsilon_{CN}$} & $1\times10^{-5}$ & $2.6\times10^{-4}$ & $5.1\times10^{-4}$  \\
 \hline
0.7 & 6 & 6 & 5\\
 \hline
0.8 & 7 & 6 & 5\\
 \hline
0.9 & 7 & 6 & 6 \\
 \hline

\end{tabular}
\end{center}
\caption{Simulated optimal repeater parameters of n for different combinations of $\eta_s$ and $\epsilon_{CN}$ in Fig. \ref{Fig6}.}
\label{Optimaln}
\end{table}

\subsubsection{Effective secret key rate}

\begin{figure}[!h]
\centerline{\includegraphics[width=0.95\columnwidth]{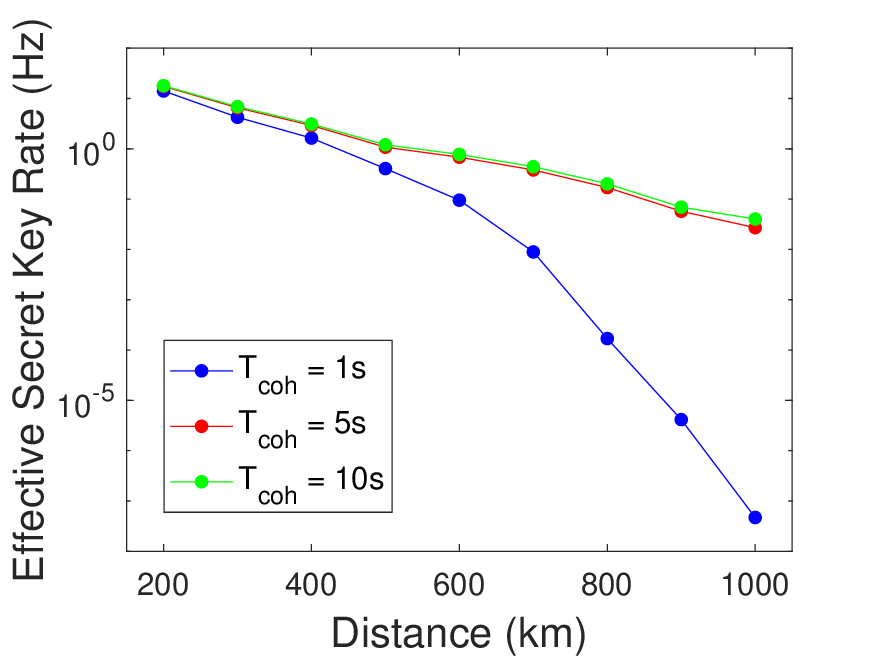}}
	\caption{(a) Simulated effective secret key rate for communication distance between 200km and 1000km with the parameters listed at the Table. \ref{Parameters}, $n_m$ = 10.}
	\label{Fig11}
\end{figure}

There is another parameter to assess the repeater's performance,
\begin{equation}
    R_{eff} = R\frac{1}{2^nn_m}\frac{L}{L_{att}},
\end{equation}
which is the effective secret key rate per unit resource usage and attenuation length \cite{muralidharan2014ultrafast, borregaard2020one}. The simulated results are shown in Fig. \ref{Fig11}.

\subsection{Entanglement purification}

\begin{figure}[!h]
\centerline{\includegraphics[width=0.95\columnwidth]{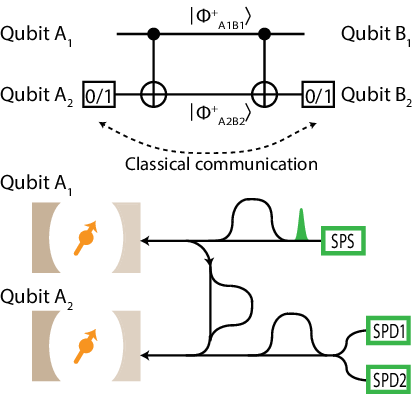}}
	\caption{(a) Entanglement purification: two qubits $A_1$ and $B_1$ are entangled as $\ket{\Phi_{A_1B_1}^+}$; two qubits $A_2$ and $B_2$ are entangled as $\ket{\Phi_{A_2B_2}^+}$; CNOT gate operation is performed between qubits $A_1$ and $A_2$; CNOT gate operation is performed between qubits $B_1$ and $B_2$; State measurements are performed to qubits $A_2$ and $B_2$; Measurement results are sent to each other to decide to keep or discard the pair $\ket{\Phi_{A_1B_1}^+}$.}
	\label{FigEP}
\end{figure}

Two entangled pairs can be used to perform two CNOT gate operation between each of the pairs, followed by state detection of one pair and communication of the measurement results to each other, which can boost the fidelity of the other pair. CNOT gate can be implemented by exploiting the strong interactions between two single atoms placed close to each other. Here I introduce another scheme of CNOT gate between two distant single atoms coupled with optical cavities mediated by one single photon, shown in Fig. \ref{FigEP}. First single photon passes through an UMZI to be prepared with the state $\ket{+}_{p}$ = 1/$\sqrt{2}$($\ket{e}$ $+$ $\ket{l}$). Subsequently the photon is sent to the first cavity and get reflected, then sent through another UMZI, which converts $\ket{+}_{p}$ = 1/$\sqrt{2}$($\ket{e}$ $+$ $\ket{l}$) to $\ket{l}$, and $\ket{-}_{p}$ = 1/$\sqrt{2}$($\ket{e}$ $-$ $\ket{l}$) to $\ket{e}$. After that, the photon is sent to the second cavity and get reflected, followed by photon measurement in the basis $\ket{\pm}$, and the measurement result is sent to the second atomic system, unitary operations to the atom maybe needed based on the measurement result. The operation described above constitute a CNOT gate between the two single atom, with the first atom in $\ket{0/1}$ basis and the second atom in $\ket{\pm}_{a}$ basis,
\begin{equation}
\begin{split}
     \ket{0}\ket{+} \rightarrow \ket{0}\ket{+}; \\
     \ket{1}\ket{+} \rightarrow \ket{1}\ket{-}; \\
     \ket{0}\ket{-} \rightarrow \ket{0}\ket{-}; \\
     \ket{1}\ket{-} \rightarrow \ket{1}\ket{+}.
\end{split}
\end{equation}
If one hadamard gate is applied to the second atom before the operations and another Hadamard gate is applied to the same atom after the operations, a CNOT gate between the two atoms is implemented with both atoms in basis $\ket{0/1}$. The detailed calculation of the protocol is shown at the Appendix. \ref{CNOTC}. If the two atoms are close to each other such as in the same cavity, Rydberg blockade can be explored to perform CNOT gate operation between them \cite{urban2009observation, welte2018photon}. 
As described earlier, non-local and non-destructive BSM can be applied to distant two atoms, so quantum Zeno effect can be used to prevent the decoherence of the entangled pairs \cite{misra1977zeno, facchi2008quantum}. Constant BSM will project the state to their original Bell state, which prevent detrimental effects to the atoms by freezing the quantum state evolution over time.

\section{Discussions}
The qubit state can be swapped to a neighboring nuclear spin with sufficient long coherence time, which has been demonstrated with NV canter, SiV center, $\prescript{1}{}{\text{H}}$ nuclear spin with erbium ion in CaWO$_4$ \cite{uysal2023coherent}. Hybrid atomic systems, such as one atom acts as communication qubit and the other atom function as memroy qubit \cite{drmota2023robust}. In this way, errors created by decoherence can be significantly reduced with the help of high-fidelity swapping gate. The current fidelity for photon-atom gate is relatively low with attenuated laser pulse serving as single photon source. The fidelity can be boosted by adapting true single photon source, such as quantum dot. In this repeater scheme, photon wavepacket can be as long as 10s of microseconds, which can greatly increase the photon-atom gate fidelity, since $\kappa T \gg 1$. Unlike other quantum repeater schemes, this scheme can be implemented with various types of atomic systems \cite{dreau2018quantum,bock2018high, bersin2024telecom, liu2024creation}. Rare-earth-ion in solids have extremely long coherence time, but suffering from low decay rate, and this repeater scheme doesn't require the atoms to decay. In addition, erbium ions have optical transition wavelength in telecomm C-band. Integrating SiV centers, rare-earth ions with nanophotonic cavities, so multiplexing with a large number is readily to be achieved. Atom array with individual control of each atom have been demonstrated experimentally, so multiplexed operation with atom array can be effectively implemented \cite{huie2021multiplexed, young2022architecture, covey2023quantum}. Entangled photon pairs have been demonstrated with rates of MHz, and the rates will be in the order of 100Hz after direct transmission of 200km \cite{craddock2024high, mueller2024high}. With improved fidelity of photon-atom CNOT gate to be 99.9\% and multiplexing of 5-10 single atoms in cavities, this repeater scheme can be used to achieve higher rates compared with direct transmission in the near future, which paves the way to demonstrate quantum supremacy of quantum-repeater enhanced entanglement distribution. 

\section{Conclusion}
In conclusion, I propose a robust quantum repeater scheme which can be implemented with various types of atomic system in near future, and high secret key rates can be achieved. 

\section{Acknowledgments}
 Y.L. acknowledge the support from Northwestern University. Discussions with Dr. Christoph Simon and Xinchao Zhou are very helpful.

\section{Disclosures}
The authors declare no conflicts of interest.

\section{Data availability}
Data underlying the results presented in this paper are not publicly available at this time but may be obtained from the authors upon reasonable request.

\appendix

\section{Alternative experimental configurations for photon-atom CNOT gate}
\label{CNOTO}
\begin{figure}[!h]
\centerline{\includegraphics[width=0.95\columnwidth]{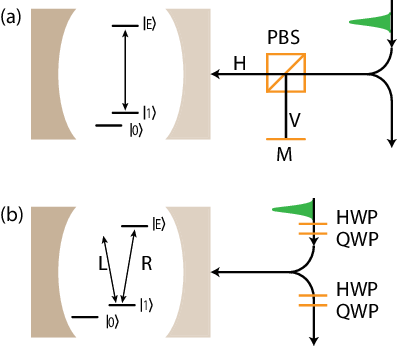}}
	\caption{(a) Photon-atom gate with linearly polarized photon and one atom in cavity \cite{duan2004scalable}. (b) Photon-atom gate with circularly polarized photon and one atom in cavity \cite{reiserer2014quantum}.}
	\label{FigA1}
\end{figure}

There are  a few different schemes for photon-atom CZ gate based on different atomic level structures and different photonic encoding, as shown in Fig. \ref{FigA1}. An atomic $\Lambda$ system with two ground states and one optically excited state. The photon is encoded with horizontal and vertical polarization. Vertical polarization is reflected by a mirror with $\pi$ phase shift. If the atom is in state $\ket{0}$, photon with $\ket{V}$ state will be reflected without entering the cavity and acquire $\pi$ phase shift; while the atom is in state $\ket{1}$, photon will enter the cavity and acquire no phase shift. This forms a CZ gate between the atom and photon. If changing the photon basis to be diagonal $\ket{D}$ = $\frac{1}{\sqrt{2}}$($\ket{H}$ + $\ket{V}$) and anti-diagonal $\ket{A}$ = $\frac{1}{\sqrt{2}}$($\ket{H}$ - $\ket{V}$), the process will be a CNOT gate. 
There is another scheme based on circularly polarized photons. The cavity is resonant with optical transition between states $\ket{0}$ and $\ket{E}$, which only interacts with right circular polarized photon. If the atom is in state $\ket{1}$ and photon is right circular polarized $\ket{R}$, there is no phase shift; while the atom is in state $\ket{1}$ and photon is right circular polarized $\ket{L}$ or the atom is in state $\ket{0}$, there will be $\pi$ phase shift, which forms a CZ gate. If changing the photon basis to be diagonal $\ket{D}$ = $\frac{1}{\sqrt{2}i}$(i$\ket{R}$ + $\ket{L}$) and anti-diagonal $\ket{A}$ = $\frac{1}{\sqrt{2}}$(i$\ket{H}$ + $\ket{V}$), the process will be a CNOT gate. Other schemes based on different atomic level structures have been proposed as well \cite{li2012robust, nagib2024robust}. 

\section{Alternative methods of deterministic BSM}
\label{BSMO}

\begin{figure}[!h]
\centerline{\includegraphics[width=0.9\columnwidth]{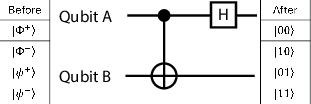}}
	\caption{Deterministic Bell state measurement with Hadamard gate and CNOT gate.}
	\label{FigA2}
\end{figure}

Direct state measurements of two qubits won't tell which Bell state they are in. For example, if the measurement result is $\ket{01}$, they can be either in state $\ket{\Psi^+}$ or $\ket{\Psi^-}$. Complete BSM can be accomplished by a single-qubit Hadamard gate and a two-qubit CNOT gate, which converts the Bell state to be a singlet state, followed by direct state measurements in basis $\ket{0}$ and $\ket{1}$, shown in Fig. \ref{FigA2}, which has been shown experimentally \cite{hermans2022qubit, kamimaki2023deterministic}.

\section{Repeater performance}
\label{RP}
\begin{figure*}[!h]
\centerline{\includegraphics[width=1.5\columnwidth]{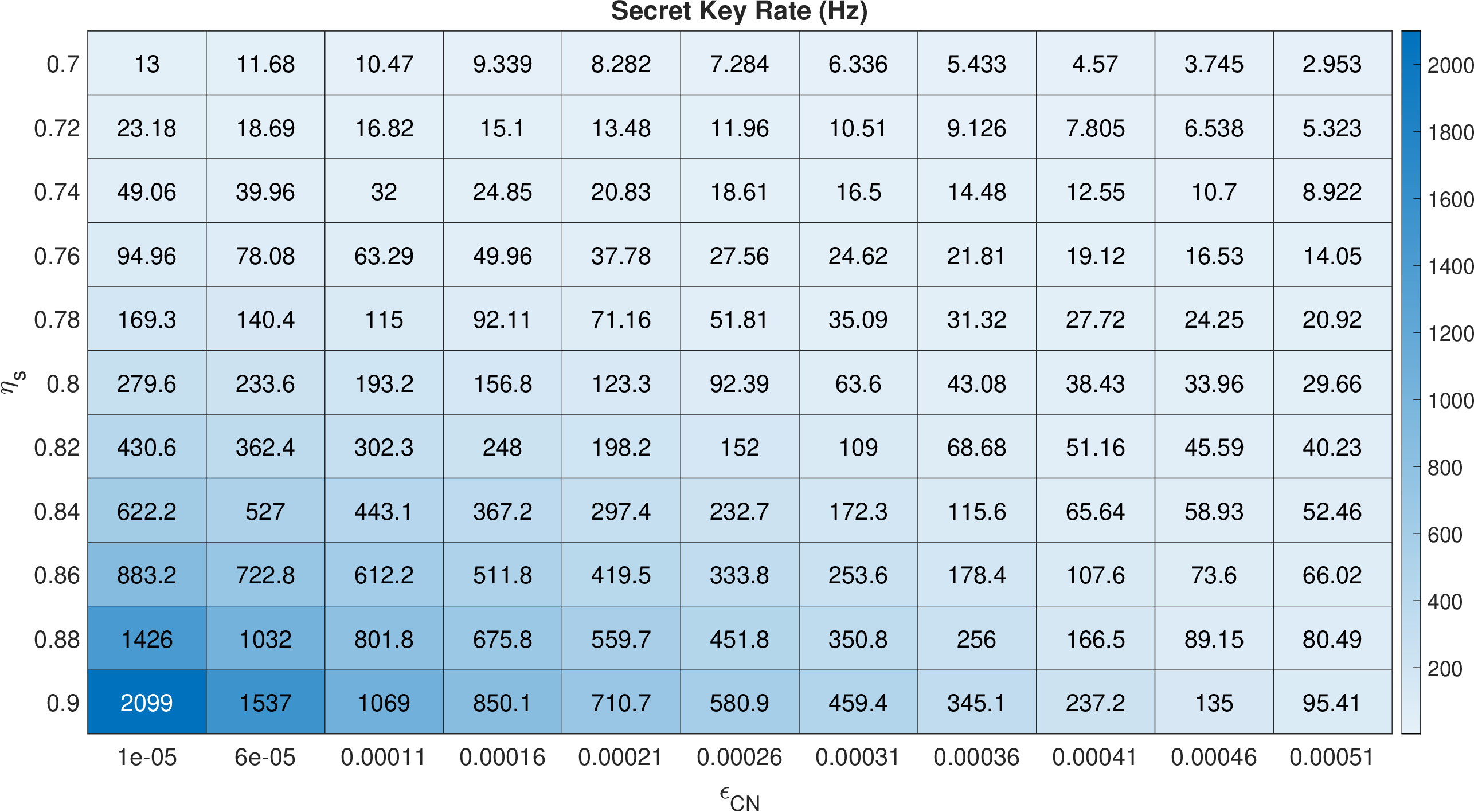}}
	\caption{(a) Simulated secret key rates for communication distance of 1000km with CNOT gate error rates ranging from $10^{-5}$ to $5.1\times10^{-4}$ and single photon source efficiency between 0.7 and 0.9, $n_m$ = 100.}
	\label{FigA3}
\end{figure*}

\begin{figure*}[!h]
\centerline{\includegraphics[width=1.5\columnwidth]{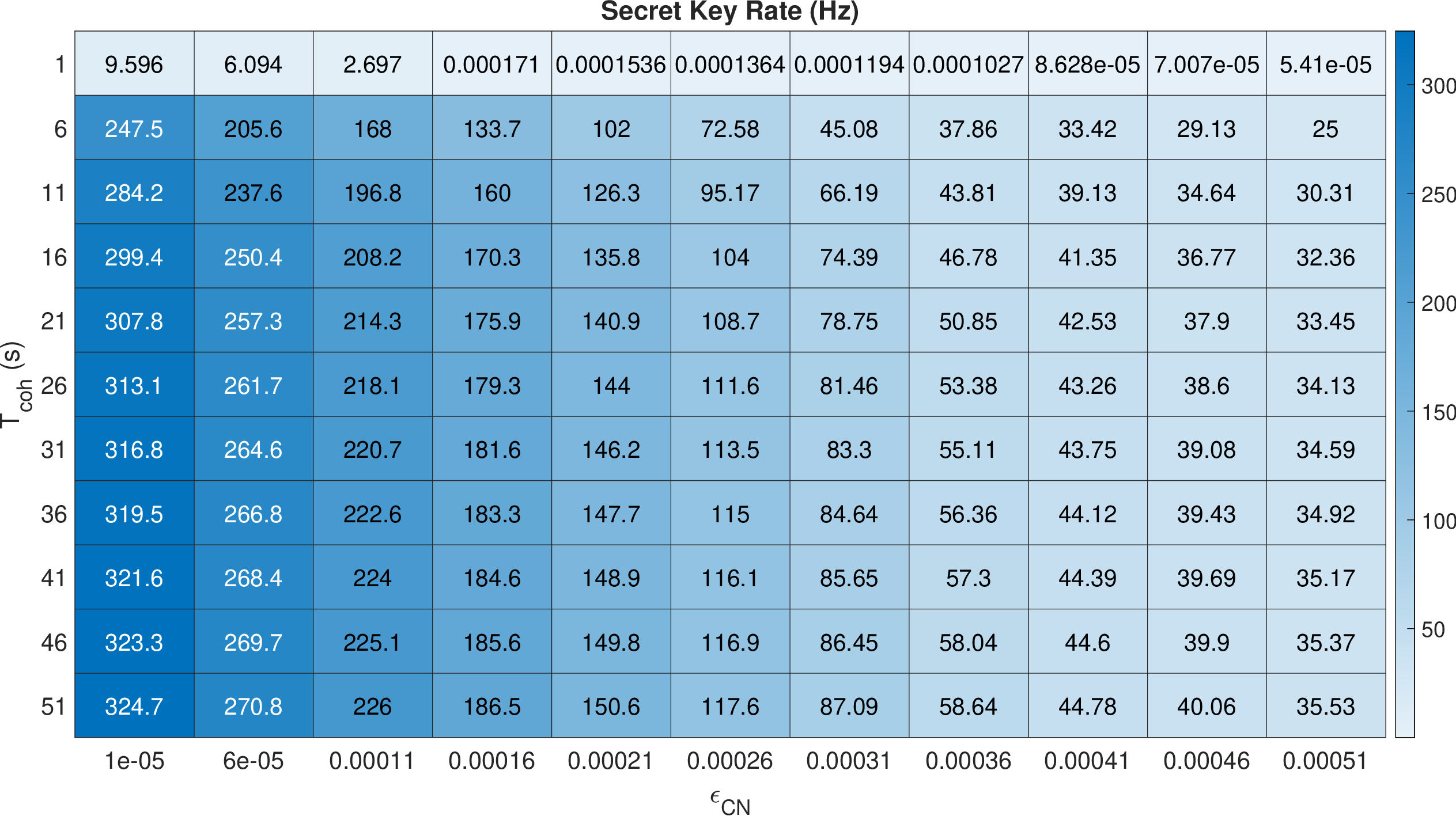}}
	\caption{Simulated secret key rates for communication distance of 1000km with CNOT gate error rates ranging from $10^{-5}$ to $5.1\times10^{-4}$ and coherence time of the qubit between 1s and 51s, $n_m$ = 100.}
	\label{FigA4}
\end{figure*}

For L = 1000km and $n_m$ = 100, Secret key rates with different combination of single photon source efficiency and CNOT gate error rate are shown in Fig. \ref{FigA3}, as well as different combination of qubit coherence time and CNOT gate error rate in Fig. \ref{FigA4}.


\section{Single-photon mediated CNOT gate between two distant atoms in cavities}
\label{CNOTC}

The mixed state of the two atoms,
\begin{equation}
    \ket{\Psi} = c_1\ket{00} + c_2\ket{01} + c_3\ket{10} + c_4\ket{11},
    \label{eqa1}
\end{equation}
and the photon is in state of $\ket{+}_{p}$ = 1/$\sqrt{2}$($\ket{e}$ $+$ $\ket{l}$), which gives,
\begin{equation}
    \ket{\Psi} = c_1\ket{00}\ket{+} + c_2\ket{01}\ket{+} + c_3\ket{10}\ket{+} + c_4\ket{11}\ket{+},
\end{equation}
after passing through the first cavity and CNOT operation between the first atom and the photon,
\begin{equation}
    \ket{\Psi} = c_1\ket{00}\ket{+} + c_2\ket{01}\ket{+} + c_3\ket{10}\ket{-} + c_4\ket{11}\ket{-},
\end{equation}
the photon passing through an UMZI,
\begin{equation}
    \ket{\Psi} = c_1\ket{00}\ket{l} + c_2\ket{01}\ket{l} + c_3\ket{10}\ket{e} + c_4\ket{11}\ket{e},
\end{equation}
the photon is sent to the second cavity and a CNOT gate operation is performed,
\begin{equation}
    \ket{\Psi} = c_1\ket{00}\ket{l} + c_2\ket{01}\ket{l} + c_3\ket{10}\ket{e} - c_4\ket{11}\ket{e},
\end{equation}
rewriting the wavefunction with the photon in $\ket{\pm}$ basis,
\begin{equation}
    \begin{split}
       \ket{\Psi} = \frac{1}{2}(\ket{e}+\ket{l})\left(c_1\ket{00} + c_2\ket{01} + c_3\ket{10} - c_4\ket{11}\right) \\
       + \frac{1}{2}(\ket{e}-\ket{l})\left(-c_1\ket{00} - c_2\ket{01} + c_3\ket{10} - c_4\ket{11}\right),
    \end{split}
\end{equation}
rewriting the wavefunction as,
\begin{equation}
    \begin{split}
       \ket{\Psi} = \frac{1}{2}(\ket{e}+\ket{l})\left(\ket{0}(c_1\ket{0} + c_2\ket{1}) + \ket{1}(c_3\ket{0} - c_4\ket{1})\right) \\
       + \frac{1}{2}(\ket{e}-\ket{l})\left(-\ket{0}(c_1\ket{0} + c_2\ket{1}) + \ket{1}(c_3\ket{0} - c_4\ket{1})\right).
    \end{split}
\end{equation}
If the photon measurement result is $\ket{+}_{p}$ = 1/$\sqrt{2}$($\ket{e}$ $+$ $\ket{l}$), the wavefunction collapse into,
\begin{equation}
    \begin{split}
       \ket{\Psi} = \frac{1}{\sqrt{2}}\left(\ket{0}(c_1\ket{0} + c_2\ket{1}) + \ket{1}(c_3\ket{0} - c_4\ket{1})\right);
    \end{split}
    \label{eqa2}
\end{equation}
If the photon measurement result is $\ket{-}_{p}$ = 1/$\sqrt{2}$($\ket{e}$ $-$ $\ket{l}$), a $\pi$ pulse around y-axis then another $\pi$ pulse around x-axis are applied to the atom, and the wavefunction is converted into,
\begin{equation}
    \begin{split}
       \ket{\Psi} = \frac{1}{\sqrt{2}}\left(\ket{0}(c_1\ket{0} + c_2\ket{1}) + \ket{1}(c_3\ket{0} - c_4\ket{1})\right).
    \end{split}
    \label{eqa2}
\end{equation}
The last two terms can be rewritten as the following:
\begin{equation}
    \begin{split}
        c_3\ket{0} + c_4\ket{1} \Leftrightarrow \frac{c_3+c_4}{2}(\ket{0} + \ket{1}) + \frac{c_3-c_4}{2}(\ket{0} - \ket{1}) \\
        c_3\ket{0} - c_4\ket{1} \Leftrightarrow \frac{c_3-c_4}{2}(\ket{0} + \ket{1}) + \frac{c_3+c_4}{2}(\ket{0} - \ket{1}),
    \end{split}
\end{equation}
comparing equations \ref{eqa1} and \ref{eqa2}, it is a CNOT gate between the two atoms, with the first atom in $\ket{0/1}$ basis and the second atom in $\ket{\pm}_{a}$ basis. If the measurement result is $\ket{}_{p}$ = 1/$\sqrt{2}$($\ket{e}$ $-$ $\ket{l}$), unitary operation is needed to convert,
\begin{equation}
    \begin{split}
    c_1\ket{00} - c_2\ket{01} + c_3\ket{10} - c_4\ket{11} \\
        \Rightarrow \ket{0}(c_1\ket{0} + c_2\ket{1}) + \ket{1}(c_3\ket{0} - c_4\ket{1}.
    \end{split}
\end{equation}

Similar calculations for circularly polarized photons are shown in this reference \cite{daiss2021quantum}.


 


\section{Two-sided cavity}
With two-sided cavity, with one photon successively reflected back by two cavities, the photon and two atoms are in state,
\begin{equation}
    \ket{p,A,B} = \frac{1}{\sqrt{2}} \left( \ket{+}\ket{\Phi^+_{AB}} + \ket{+}\ket{\Phi^-_{AB}}\right).
\end{equation}
Complete BSM with two photons reflected from both of the cavities won't work for this case, but the complete BSM method listed in Appendix. \ref{BSMO} can be used for performing complete BSM, which result in deterministic entanglement swapping.

\bibliography{sample}{}
\end{document}